\documentclass{ws-4-MG14}            
\begin{document}
\title{Sparsity of the Hawking flux}

\author{Matt Visser, Finnian Gray, Sebastian Schuster, and Alexander Van--Brunt}

\address{School of Mathematics and Statistics, Victoria University of Wellington,\\
Wellington 6140, New Zealand\\
E-mail: \{matt.visser, finnian.gray, sebastian.schuster, alexander.vanbrunt\}@msor.vuw.ac.nz\\
http://www.victoria.ac.nz/sms}



\begin{abstract}
It is (or should be) well-known that the Hawking flux that reaches spatial infinity is extremely sparse, and extremely thin,
with the Hawking quanta, one-by-one, slowly dribbling out of the black hole. 
The typical time between quanta reaching infinity is much larger than the timescale set by 
the energy of the quanta.
Among other things, this means that the Hawking evaporation of a black hole should
be viewed as a sequential cascade of 2-body decays.
\end{abstract}


\bodymatter
\def\d{{\mathrm{d}}}

\section{Introduction}

The sparsity of the Hawking flux is a 40-year-old result that has largely been forgotten.~\cite{Gray:2015, Gray:2015b, Page:1976a, Page:1976b, Page:1977, Page:thesis} Previous analyses relied  largely on numeric (rather than analytic) estimates, and focussed mainly on the late-time high-temperature regime in the final stages of the evaporation process.~\cite{Oliensis:1984, MacGibbon:1990, Halzen:1991, MacGibbon:2007, Page:2007, MacGibbon:2010}  We have developed some semi-analytic estimates to seek a deeper understanding of the underlying physics.~\cite{Gray:2015, Gray:2015b}
(See also van Putten.~\cite{van-Putten1, van-Putten2})
We assume an exact Planck spectrum, which is not the full story, but is sufficient to give tolerable estimates, certainly  for spin-zero bosons.  After carefully separating out super-radiant contributions, (which in Hod's numerically based article~\cite{Hod} were lumped in with the Hawking effect), sparsity of the Hawking flux is seen to persist throughout the entire evaporation process.

\section{Strategy}

We compare and contrast two approaches:

--- As a zeroth-order approximation we treat the Hawking flux as blackbody radiation. A  more careful treatment should at the very least include greybody, phase-space, and adiabaticity effects,~\cite{Thermality,adiabatic1,adiabatic2,essential, observability} but a blackbody approximation should be quite sufficient to set the scale for the relevant issues we wish to consider. 

--- The most significant limitation on treating the Hawking flux as  blackbody radiation comes from the greybody factors. Page~\cite{Page:1976a,Page:1976b,Page:1977,Page:thesis} resolves the Hawking flux into spin-dependent angular-momentum modes, and calculates various quantities
\begin{equation}
\langle Q \rangle =\sum_{\ell m} \int T_{s\ell m}(\omega) \; \langle n\rangle_\omega \; Q(\omega)\; {\d \omega}.
\end{equation}
Here $\langle n\rangle_\omega$ is a bosonic/fermionic occupation number, 
while the $T_{s\ell m}(\omega)$ are spin-dependent greybody factors, estimated by numerically solving the appropriate Regge--Wheeler/Zerilli equation, this all being followed by a numerical integration over frequencies. 

These two approaches give a good qualitative and quantitative handle on the sparsity of the Hawking flux. The blackbody emission approximation works best for spin-zero, with higher spins seeing extra suppression (and increased sparsity) due to the angular momentum barrier. 

\section{Flat space preliminaries}

The differential number flux, $(quanta)/(time)$, (of massless bosonic quanta emitted by a black body of temperature $T$, 
and infinitesimal surface area $\d A$),  into a wave-number range $\d k$ is (in flat space) given by:
\begin{equation}
\d \Gamma =    {g\over8\pi^2}  \;  {c k^2 \over \exp(\hbar ck/k_BT)-1 }   \; \d k \; \d A.
\end{equation}
For an object of finite surface area $A$ the total emitted number flux is:
\begin{equation}
\Gamma =      {g \,\zeta(3)\over4\pi^2}   \; {k_B^3 T^3\over\hbar^3 c^2} \; A.
\end{equation}
The reciprocal of this quantity, $\tau_\mathrm{gap} = 1/\Gamma$, 
is  the \emph{average} time interval between the emission of successive quanta. 

The peak in the number spectrum occurs where $k^2/(e^{\hbar ck/k_BT}-1)$ is maximized, 
\begin{equation}
\omega_\mathrm{\,peak\,number} = c k_\mathrm{\,peak\,number} =     {k_B T\over \hbar} \; \left( 2 + W(-2e^{-2})\right).
\end{equation}
Here $W(x)$ is the Lambert $W$-function,\cite{Lambert1, Lambert2, primes} defined by $W(x) e^{W(x)} = x$. 
Quanta emitted at this peak can only be temporally localized to within a few oscillation periods, 
so it is safe to take $\tau_\mathrm{localization} = 1/\nu_\mathrm{peak\,number}  = 2\pi/\omega_\mathrm{peak\,number}$ as a good estimate of the time required 
for each individual quantum to be emitted.
Define the dimensionless figure of merit
\begin{equation}
\eta_\mathrm{\,peak\,number}= {\tau_\mathrm{\,gap}\over\tau_\mathrm{\,localization}} =  {\nu_\mathrm{\,peak\,number}\over \Gamma } 
= {\pi\left( 2 + W(-2e^{-2})\right)\over g\; \zeta(3)} \; {\hbar^2 c^2\over k_B^2 T^2 A}.
\end{equation}
In terms of the so-called ``thermal wavelength'', $\lambda_\mathrm{\,thermal} = 2\pi \hbar c/(k_B T)$, this is
\begin{equation}
\eta_\mathrm{\,peak\,number} 
= {\left( 2 + W(-2e^{-2})\right)\over4 \pi g \;\zeta(3)} \; {\lambda^2_\mathrm{\,thermal}\over A}.
\end{equation}
If instead we consider the peak in the energy flux, 
\begin{equation}
\eta_\mathrm{\,peak\,energy} 
= {\left( 3 + W(-3e^{-3})\right)\over4 \pi g \;\zeta(3)} \; {\lambda^2_\mathrm{\,thermal}\over A}.
\end{equation}
Similarly, we could use the average frequency to set the localisation timescale
\begin{equation}
\langle \omega \rangle  = {\int ck (\d \Gamma/\d k) \d k\over  \int (\d \Gamma/\d k)  \d k } = {\pi^4\over30\;\zeta(3)}\; {k_B T\over\hbar}. 
\end{equation}
In this case
\begin{equation}
\eta_\mathrm{\,average\,energy} 
= {\pi^2\over120 g \;\zeta(3)^2} \; {\lambda^2_\mathrm{\,thermal}\over A}.
\end{equation}
More subtly, divide the spectrum into ``wave-number bins'' and  define
\begin{equation}
\eta_\mathrm{\,binned} =  {1\over \displaystyle \int {2\pi\over c k} \; {\d\Gamma\over\d k}\;  \d k}.
\end{equation}
A brief calculation yields
\begin{equation}
\eta_\mathrm{\,binned} = {24\over 2\pi g} \;{\hbar^2 c^2 \over k_B^2 T^2 A} =  {24\over 8\pi^3 g} {\lambda^2_\mathrm{\,thermal}\over A}.
\end{equation}
All of these sparsity estimates (in flat Minkowski space for now) take the form
\begin{equation}
\eta = \hbox{(dimensionless number)} \; {\lambda^2_\mathrm{\,thermal}\over g \, A}.
\end{equation}
Let us now introduce key aspects of black hole physics, adapting the discussion above to see how far we can get.

\section{Non-super-radiant situations}

Under normal laboratory (and astronomical) conditions one is dealing with emitters whose surface area is extremely large in terms of the thermal wavelength, so in those situations  $\eta \ll 1$. 
However, this is exactly what fails for a Schwarzschild black hole.
For the Hawking temperature we have
\begin{equation}
k_B T_H =  {\hbar c\over 4\pi \; r_H}; \qquad \lambda_\mathrm{\,thermal} =  8 \pi^2 \; r_H.
\end{equation}
The thermal wavelength is $8\pi^2 \approx 78.95 \approx 80$ times larger than the Schwarzschild radius.
At high frequencies (the ray optics limit)~\cite{Page:1976a} the cross section is universally given by ${27\over4} \pi r_H^2$. This implies  $A\to A_\mathrm{\,effective} =  {27\over4} A_H = 27 \pi r_H^2$.

With these substitutions, for a Schwarzschild black hole we have
\begin{equation}
\eta_\mathrm{\,peak\, number} =  {32\pi^2\left( 2 + W(-2e^{-2})\right)\over27g\zeta(3)} = {15.50768123...\over g} \gg 1.
\end{equation}
As promised, the gap between successive Hawking quanta is on average much larger than the natural timescale associated with individual emitted quanta. 
Similar calculations apply for the  other options we had considered for the localization timescale. Still working with the Schwarzschild black hole, we see that:

\noindent
--- If we consider the peak energy flux, rather than the peak number flux, then
\begin{equation}
\eta_\mathrm{\,peak\, energy} =  {32\pi^2\left( 3 + W(-3e^{-3})\right)\over27g\zeta(3)} = {27.45564528...\over g} \gg 1.
\end{equation}

\noindent 
--- If we consider the average frequency then 
\begin{equation}
\eta_\mathrm{\,average\,energy} = {26.28537289...\over g}.
\end{equation}

\noindent
--- For the binned version of the $\eta$ parameter we have 
\begin{equation}
\eta_\mathrm{\,binned}= {14.22222222...\over g}.
\end{equation}
Whatever the precise definition of $\eta$, 
it is clear that the time interval between successive emitted Hawking quanta, is on average, 
large compared to the natural timescale associated with the energy of the individual emitted quanta. 

We then compared this analysis~\cite{Gray:2015} with numerical estimates along the lines of Page's results from the mid 1970's.~\cite{Page:1976a,Page:1976b,Page:1977,Page:thesis}  The semi-analytic estimates work best for spin-zero, with higher spins being increasingly suppressed by the angular momentum barrier. The semi-analytic estimates seem to universally act as a lower bound on the value of $\eta$ extracted by numerical means.~\cite{Gray:2015, Gray:2015b}

We have performed similar semi-analytic estimates for fermions and Boltzmann particles, 
for neutral particles emitted by a Reissner--Nordstr\"om black hole, for particles emitted by dirty black holes~\cite{dbh} surrounded by matter which satisfies the weak and null energy conditions (WEC and NEC)~\cite{twilight, Visser:1999}, and for massive particles.~\cite{Gray:2015}  In the absence of super-radiance the semi-analytic results give good qualitative understanding of the underlying physics.~\cite{Gray:2015, Gray:2015b}

\section{Super-radiant situations}

Super radiance can occur when for one reason or another both the particle occupation number and the greybody factor simultaneously become negative. This can occur either because of electric charge (of \emph{both} the black hole and the emitted quantum) or because of angular momentum.~\cite{Boonserm}
Concentrate on angular momentum.
In the extremal limit ($\kappa\to0$) the greybody factors are approximately~\cite{Page:1976a}
\begin{equation}
T_{\ell m}(\omega) \approx C_{\ell, s} \; (A_H\; \omega \;[\omega-m\Omega_H])^{2\ell+1}.
\end{equation}
The bosonic occupation number is
\begin{equation}
\langle n\rangle_\omega = {1\over \exp\{\hbar(\omega-m\Omega_H)/k_B T_H\} -1 }.
\end{equation}
These both change sign at $\omega=m\Omega_H$,
and for  $0 \leq \omega \leq m \Omega_H$ the super-radiant emission is not well-approximated by a blackbody.
So it makes sense to split the integral into two regions:
\begin{itemize}
\item $0 \leq \omega \leq m \Omega_H$ \;\;--- super-radiant emission.
\item $m \Omega_H \leq\omega\leq\infty$ --- Hawking emission.
\end{itemize}
Based on a numerical analysis, it is well-known that  in the extremal limit ($\kappa\to0$) super-radiance dominates over the Hawking flux.~\cite{Page:1976a,Page:1976b,Page:1977,Page:thesis}
Semi-analytically, concentrate on the binned version of $\eta$, then
\begin{equation}
{1\over\eta} =2\pi \sum_{\ell m} \int T_{\ell m}(\omega) \; \langle n\rangle_\omega \; {\d \omega \over\omega},
\end{equation}
and in the near-extremal limit:
\begin{equation}
{1\over\eta} 
\approx 
(A_H \; \Omega_H^2)^{2\ell+1} 
\sum_{\ell m}  m^2 \;C_{\ell, s} \int \;  {(x[x-1])^{2\ell+1}  \over \exp(\epsilon[x-1])-  1  }\; {\d x\over x} .
\end{equation}
Here:
 $x = \omega/(m \Omega_H)$ and 
$\epsilon = (\hbar m \Omega_H)/(k_B T_H) \gg\!\!\gg 1$. 
As usual, the emission is dominated by the lowest available angular momentum state $\ell=m=s$:
\begin{equation}
{1\over\eta} \approx C_{s, s} \; (A_H \; s^2 \; \Omega_H^2)^{2s+1}  \int \;  {(x[x-1])^{2s+1}  \over \exp(\epsilon[x-1])-  1  }\;  {\d x\over x} .
\end{equation}
Since normal Hawking radiation and super-radiance are alternate decay channels which take place  simultaneously, it is sensible to split ``in parallel'':
\begin{equation}
{1\over\eta} = {1\over\eta_\mathrm{\,super\hbox{-}radiant}}  + {1\over\eta_\mathrm{\,Hawking}}.
\end{equation}
We have
\begin{equation}
{1\over\eta_\mathrm{\,super\hbox{-}radiant}} \approx C_{s, s} \; (A_H s^2 \Omega_H^2)^{2s+1} \int_0^1 \;  {(x[x-1])^{2s+1}  \over \exp(\epsilon[x-1])-  1  }\;  {\d x\over x} ;
\end{equation}
and
\begin{equation}
{1\over\eta_\mathrm{\,Hawking}} \approx C_{s, s}\; (A_H s^2 \Omega_H^2)^{2s+1}  \int_1^\infty \;  {(x[x-1])^{2s+1}  \over \exp(\epsilon[x-1])-  1  }\;  {\d x\over x} .
\end{equation}
Recalling that $\epsilon  \gg\!\!\gg 1$,
some brute-force integration yields the analytic estimates:
\def\O{{\mathcal{O}}}
\begin{equation}
{\eta_\mathrm{\,super\hbox{-}radiant}}  = \O\left(\epsilon^0\right); \qquad 
{\eta_\mathrm{\,Hawking}} = \O\left(\epsilon^{2s+2}\right) \gg\!\!\gg 1.
\end{equation}
So, now based on a semi-analytic analysis,  we see that super-radiance dominates in the extremal limit.
In fact, super-radiance leads to rapid spin-down with small energy loss,~\cite{Page:1976a, Page:1976b, Page:1977, Page:thesis} 
until the system goes non-super-radiant, and then the ``normal'' Hawking effect takes over.
This explains Hod's numerically based results.~\cite{Hod}
The quantitative details are messy, but the overall message is clear: Sparsity of the Hawking flux is the dominant feature of the Hawking evaporation process.

\enlargethispage{15pt}
\section{Discussion}

For non-super-radiant modes the Hawking flux is extremely sparse --- the average time between emission of Hawking quanta is very large compared to the timescale set by the energies of the Hawking quanta.  The Hawking quanta are dribbling out one-by-one, with very large interstitial gaps.  This phenomenon persists throughout the entire evolution of the black hole, both in early stages and (modulo super-radiance) in late stages. 
Compared to numerics the semi-analytic estimates often under-estimate sparsity by factors of 100 or even more.
Sparsity, is here to stay --- modulo technical and linguistic arguments on how to classify super-radiance. 

\noindent
The sparseness of the Hawking flux has a number of kinematical implications:
\begin{itemlist}
\item While early-stage Hawking radiation from Schwarzschild or Reissner--Nordstr\"om black holes is spherically symmetric, this spherical symmetry is only a long-term statistical statement obtained after averaging over very many Hawking quanta. 
\item
Early-stage Hawking evaporation should be seen as a long chain of independent 2-body decay processes involving photons, gravitons, and neutrinos.~\cite{Visser:1992}  (Similarly, late-stage Hawking evaporation, once the temperature exceeds $\Lambda_\mathrm{\,QCD}$, should be viewed as a long chain of 2-body decay processes proceeding via the emission of hadronic jets.~\cite{Oliensis:1984, MacGibbon:1990, Halzen:1991, MacGibbon:2007, Page:2007, MacGibbon:2010})
\item
When analyzing the emission of individual Hawking quanta one should use the special relativistic kinematics that is applicable in the asymptotic spatial region.~\cite{Visser:1992} 
Depending on whether or not one views black hole masses as being quantized or continuous, one can view this either as a normal 2-body decay, or as the decay of one IMP (``indefinite mass particle'' $\approx$ unparticle) into another IMP.~\cite{IMP, IMP2} 
It may be profitable to reconsider and reanalyze the entire Hawking evaporation process from this point of view.
\end{itemlist}




\end{document}